\documentclass[angeo]{copernicus}

\newcounter{IonCS}

\newcounter{RomC}

\newcommand{\be}{\begin{equation}}
\newcommand{\ee}{\end{equation}}
\newcommand{\ba}{\begin{eqnarray}}
\newcommand{\ea}{\end{eqnarray}}
\usepackage{amsmath}

\begin{document}

\title{Topologically Driven Coronal Dynamics - A Mechanism for Coronal Hole Jets}

\author[1]{D.A.N. M{\"u}ller}
\author[2]{S.K. Antiochos}

\affil[1]{European Space Agency, Research and Scientific Support Department\\ c/o NASA Goddard Space Flight Center, Mail Code 671.1, Greenbelt, MD 20771, USA}
\affil[2]{Heliophysics Division, NASA Goddard Space Flight Center, Mail Code 674.0, Greenbelt, MD 20771, USA}

\runningtitle{A Mechanism for Coronal Hole Jets}

\runningauthor{D.A.N. M{\"u}ller and S.K. Antiochos}

\correspondence{D.A.N. M{\"u}ller\\ (dmueller@esa.nascom.nasa.gov)}

\received{}
\pubdiscuss{} %% only important for two-stage journals
\revised{}
\accepted{}
\published{}

\firstpage{1}

\maketitle

\begin{abstract}
Bald patches are magnetic topologies in which the magnetic field is
concave up over part of a photospheric polarity inversion line. A bald
patch topology is believed to be the essential ingredient for filament
channels and is often found in extrapolations of the observed
photospheric field. Using an analytic source-surface model to calculate the magnetic topology of a small bipolar region embedded in a global magnetic dipole field, we demonstrate that although common in closed-field regions close to the solar equator, bald patches are unlikely to occur in the open-field topology of a coronal hole. 
Our results give rise to the following question: What happens to a bald patch topology when the surrounding field lines open up? This would be the case when a bald patch moves into a coronal hole, or when a coronal hole forms in an area that encompasses a bald patch. Our magnetostatic models show that, in this case, the bald patch topology almost invariably transforms into a null point topology with
a spine and a fan. We argue that the time-dependent evolution of this
scenario will be very dynamic since the change from a bald patch to
null point topology cannot occur via a simple ideal evolution in the
corona.  We discuss the implications of these findings for recent
Hinode XRT observations of coronal hole jets and give an outline of
planned time-dependent 3D MHD simulations to fully assess this
scenario.
\end{abstract}

\introduction

Magnetic bald patches are sections of a polarity inversion line (PIL) where the field lines are concave up rather than concave down as is the usual case for a loop geometry (a PIL is any location on the photosphere where the normal flux changes sign).
The magnetic field lines forming a bald patch belong to a separatrix surface along which a current sheet may be formed by shearing motions of magnetic footpoints at the photosphere \citep{Low1987ApJ,Low+Wolfson1988ApJ,Wolfson1989ApJ,Vekstein+al1991AA,Vekstein+Priest1992ApJ,Karpen+al1990ApJ}.
Conditions for the appearance of bald patches at the solar surface were first described by \citet{Titov+al1993AA} and then extended by \citet{Bungey+al1996AA}.
A common flux distribution that leads to bald patch formation is that of a small opposite polarity region embedded in a large scale unipolar background field. 
Such a nested polarity will appear if, for example, a bipolar active region emerges in a unipolar hemisphere, or if an ephemeral region emerges in a coronal hole or if a small magnetic carpet bipole emerges in the unipolar network. Consequently, we expect this polarity distribution to be a generic feature of the solar photosphere.

\begin{figure}[ht]
\vspace{2mm}
\begin{center}
\includegraphics[width=0.4\textwidth,angle=-90]{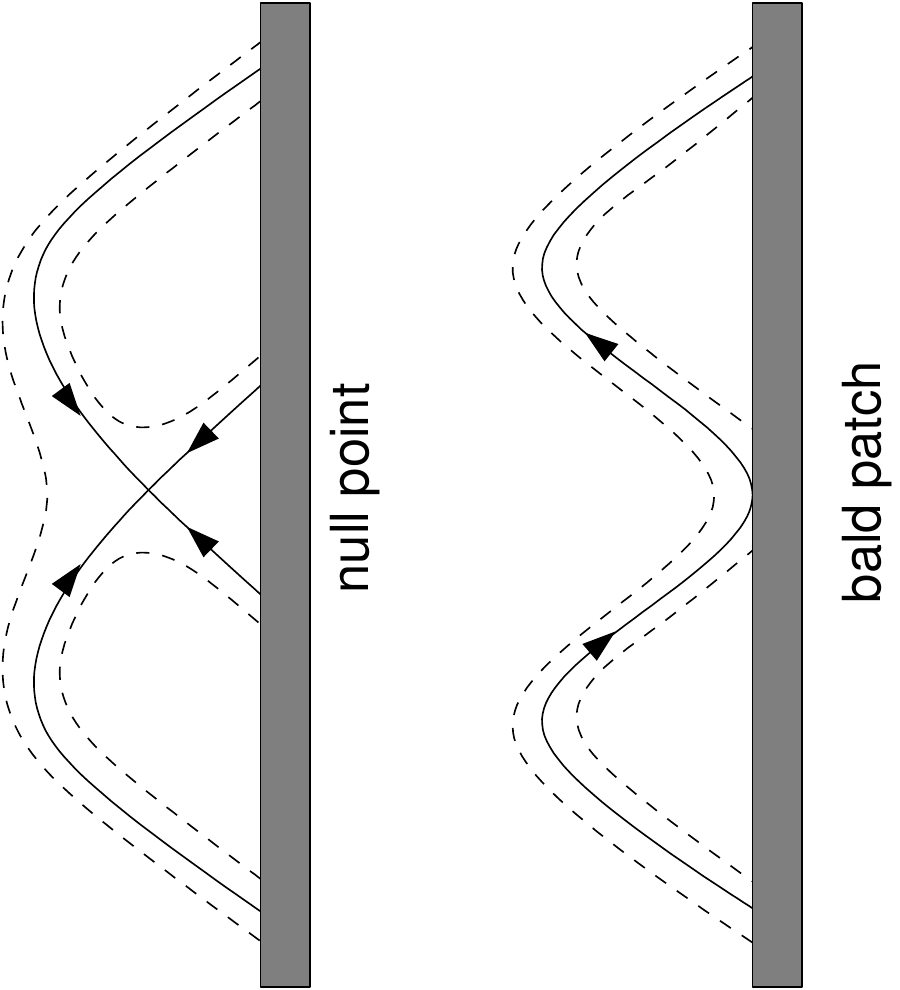}
\end{center}
\caption{\label{fig:bp_null_sketch}Sketch of two different magnetic topologies: Null point (\emph{top}) and bald patch (\emph{bottom}).}
\end{figure}

\begin{figure*}[ht]
\vspace*{2mm}
\begin{center}
\includegraphics[width=0.48\textwidth]{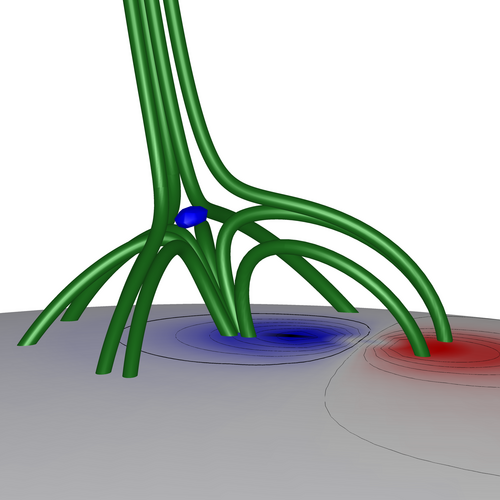}
\includegraphics[width=0.48\textwidth]{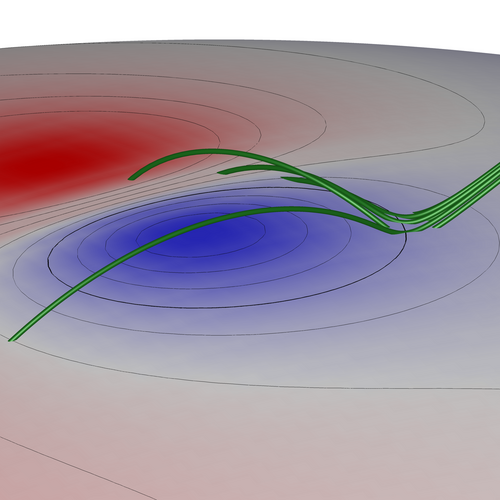}
\end{center}
\caption{\label{fig:n_bp}A magnetic null point (\emph{left}) and a bald patch topology (\emph{right}). Red shading indicates positive, blue shading negative polarity of the radial magnetic field strength, $B_r$. The thick black lines indicate the polarity inversion lines (PILs), the blue bubble marks the location of the magnetic null point, and some magnetic field lines are drawn in green. For a bald patch topology, the magnetic null lies below the surface, and the field lines are parallel to the surface and concave up along part of the PIL.}
\end{figure*}

There are only two possible magnetic topologies for such a nested polarity region: Null points in the corona and magnetic bald patches.
Figure~\ref{fig:bp_null_sketch} shows a sketch of these two topologies, while Fig.\ \ref{fig:n_bp} shows a 3D rendering of field lines for both cases.
%f
Note that as long as there is only one nested polarity region, i.e.\ only one closed PIL in a large scale unipolar photospheric region, then the statements above must hold independently of the details of the distribution of the flux \citep{Antiochos+al2007ApJ}.

The best known and most widely studied topology of the two is that of a null point in the corona, along with the usual dome-shaped fan separatrix surface and pair of spine lines \citep[e.g.][]{Greene1988JGR,Lau+Finn1990ApJ,Antiochos1990MmSAI,Priest+Titov1996Phil}.
On the other hand, a bald patch occurs if a null point is located below the surface such that it produces concave up field lines along part of the PIL. Determining the evolution of bald patch topologies is difficult, because simple line-tied boundary conditions cannot be used wherever the coronal field is concave up at the photosphere \citep{Antiochos1990MmSAI,Karpen+al1990ApJ}.
At the same time, bald patches are highly interesting since they are probably linked to jetting activity such as surges and sprays \citep{Georgoulis+al2002ApJ}. A bald patch topology was also found to be involved in surges
by \cite{Mandrini+al2002AA}, in a small flare \citep{Aulanier+al1998SP}, and in transition region brightenings \citep{Fletcher+al2001SP}.
The primary goals of this paper are (a) to find out under which conditions magnetic bald patches exist, and (b) under which conditions existing bald patches are forced to vanish due to a change in the surrounding magnetic field.

We will describe our model and the methods in Sections~\ref{model} and \ref{method}, present some results in Section~\ref{results}, establish a relation to recent Hinode XRT observations in Section~\ref{obs}, and finally discuss our findings in Section~\ref{discussion}.

\section{\label{model}A simple magnetostatic model}
Our model consists of a large global magnetic dipole and a small local embedded dipole  which is placed slightly below the solar surface. The advantage of this simple configuration is that its magnetic potential, $\Phi$, and field, $B = - \nabla \Phi$, can be calculated analytically. For a magnetic dipole with dipole moment $\vec{d}$ at position $\vec{r}_d$ and a source surface at $r = R_s$,  the magnetic potential, $\Phi$, at a location $\vec{r}$ is given by \citep{Antiochos+al2007ApJ}
\be
\Phi = \frac{\vec{d}\cdot(\vec{r}-\vec{r}_d)}{\bigl|\vec{r} - \vec{r}_d\bigr|^3} - \frac{R_s r_d^3 \vec{d} \cdot\bigl(R_s^2 \vec{r} - r^2\vec{r}_d\bigr)}{\Bigl|r_d^2 \vec{r} - R_s^2 \vec{r}_d\bigr|^3}\, .
\label{eq:phi}
\ee

\noindent
One can easily verify that $\Phi = 0$ at the source surface, $r = R_s$, i.e.\ the field is purely radial there.
For the simplest case of a central dipole, Equation~(\ref{eq:phi}) reduces to
\be
\Phi_0  = \frac{(\vec{d}\cdot\vec{r})(R_s^3 - r^3)}{R_s^3 r^3} \,.
\ee
\noindent
The potential of a normalized central dipole can also be written as
\be
\Phi_0 = \frac{1}{2 R_s^3+1}\cdot r\cos(\theta)\cdot \Bigl(\Bigl(\frac{R_s}{r}\Bigr)^3 -1\Bigr) \,.
\ee

\noindent
The radial component of the magnetic field at the solar surface ($r=r_\odot=1$) is then simply $B_r (\theta)=\cos(\theta)$. In this expression, $\theta$ denotes the co-latitude which is zero at the pole and $\theta=\pi/2$ at the equator.

If a dipole at a location $\vec{r}_d$ is perpendicular to the radius vector, $\vec{d} \perp \vec{r}_d$, Equation (\ref{eq:phi}) can be written as
\be
\Phi_1 (\vec{r})= \frac{\vec{d}\cdot\vec{r}}{\bigl|\vec{r} - \vec{r}_d\bigr|^3} - \frac{\bigl(\frac{R_s}{r_d}\bigr)^3\vec{d}\cdot \vec{r}}{\bigl|\vec{r} - \bigl(\frac{R_s}{r_d}\bigr)^2 \vec{r}_d\bigr|^3}\, .
\ee

\begin{figure*}[ht]
\vspace*{2mm}
\begin{center}
\includegraphics[width=0.48\textwidth]{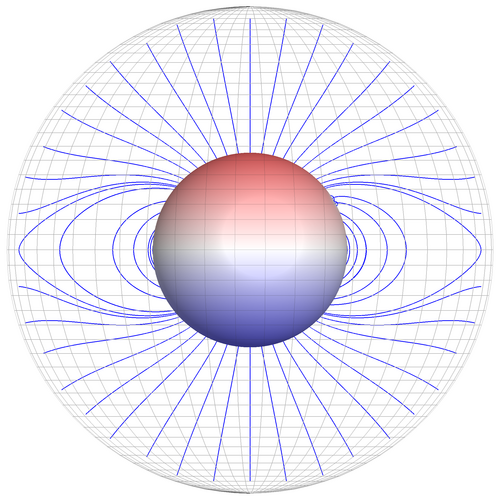}
\includegraphics[width=0.48\textwidth]{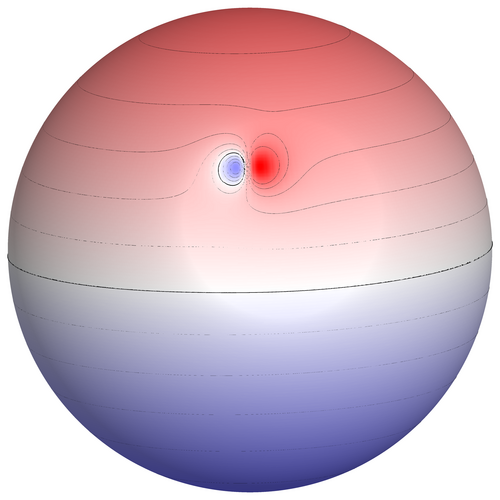}
\end{center}
\caption{\label{fig:ssg}Source surface model of the solar magnetic field. \emph{Left panel}: The field lines outline the global dipole field structure, and the source surface is indicated by a gray wire mesh. The color shading of the solar sphere corresponds to the radial magnetic field strength. The magnetic bald patch shown in the right panel of Fig.~\ref{fig:n_bp} is located at the east limb around $30^\circ$ northern latitude. \emph {Right panel}: Same magnetic configuration, this time showing the bipolar structure that gives rise to the bald patch. The thin black lines represent iso-contours of the radial magnetic field, the thick black lines are the polarity inversion lines (PIL, $B_r =0$).}
\end{figure*}

\noindent
The first term is the magnetic potential of the dipole, the second one is the potential of the corresponding image dipole. The image dipole is located at  $\vec{r}_{\rm im} = \bigl(\frac{R_s}{r_d}\bigr)^2 \vec{r}_d$ and has a dipole moment of  $\vec{d}_{\rm im} = - \bigl(\frac{R_s}{r_d}\bigr)^3 \vec{d}$. 
In this work, we consider a magnetic potential that is the sum of the potentials of a central dipole and a local dipole close to the surface, i.e.
\be
\Phi = \Phi_0 + \Phi_1 \,.
\ee
\noindent
The addition of the local dipole results in a new PIL in addition to the one at the equator, which separates the nested polarity from its surroundings.
Figure~\ref{fig:ssg} illustrates the resulting large scale magnetic field for a source surface radius of $R_s = 2.5~r_\odot$ (this value is used throughout this work). The field lines outline the global dipole field structure, and the source surface is indicated by a gray wire mesh. The color shading of the solar sphere corresponds to the radial magnetic field strength, $B_r$.

We emphasize again that although the particular fields shown in Figs.\ \ref{fig:n_bp} and  \ref{fig:ssg} were derived for two dipole sources, the magnetic topology corresponds to any flux distribution that produces a global PIL and a small nested PIL.

\section{\label{method}Determining the separatrix between null point and bald patch solutions}

As stated before, the primary goals of this study are to find out under which conditions magnetic bald patches exist, and under which conditions existing bald patches are forced to vanish due to a change in the surrounding magnetic field. As a first step, we constrain the problem by studying the conditions under which an embedded dipole supports a bald patch topology (which has a null point below the surface), and when there will be a null point above the surface. 

To address this question, we carried out a parameter study by moving an embedded dipole via a sequence of equilibrium states from close to the solar equator to the pole. Its radial position $r_d=0.94 ~r_\odot$ and its orientation in east-west direction parallel to the surface was kept constant. We then determined the critical dipole strength, $l_{\rm crit}$, that is needed to place the magnetic null right at the solar surface for all solar latitudes for which a closed patch of opposite polarity flux exists.\footnote{Depending on the exact dipole strength and radial position, the PIL associated with the embedded dipole merges with the one that separates the two solar hemispheres around $15^\circ$ latitude.} In general, for weak dipole strengths the null point will be submerged, while for larger values of $l$, the null point will be located above the surface. The dipole strength for which the magnetic null is exactly located at the solar surface will thus separate these two regimes and help to assess the physical conditions associated with each of them.

For a given latitude, this critical value of the embedded dipole strength is calculated by first determining the location, ${\vec r}_{\rm min}$, of the minimum of $B^2$ inside the discretized volume $\vec{B}(\vec{r},l)$, and then minimizing the radial distance between the location of this minimum and the solar radius, $r_\odot$, i.e.\ finding the root of 

\be
f(B^2(\vec{r},l)) = \underset{r}{\operatorname{argmin}}\, (B^2(\vec{r},l)) - r_\odot\,.
\ee

\noindent
Since $f$ is an implicit function of the dipole strength, $l$, this minimization is achieved using a modified Newton-Raphson method. In the expression
\be
l_{n+1} = l_n - \alpha\frac{f(l_n)}{f^\prime(l_n)}
\ee
of the conventional Newton-Raphson method, the analytical derivative is replaced by a numerical two-point derivative, and a convergence factor $\alpha$ is included and set to $0.5$ (instead of the usual $\alpha=1$) to avoid running into possible neighboring local minima of $f$. This as well as sufficiently high spatial resolution inside the computational domain are particularly important for a correct determination of the location of the null point since the global magnetic field decreases radially as $r^{-3}$.

\section{\label{results}Results}
\begin{figure}[t]
\vspace*{2mm}
\begin{center}
\includegraphics[width=8.3cm]{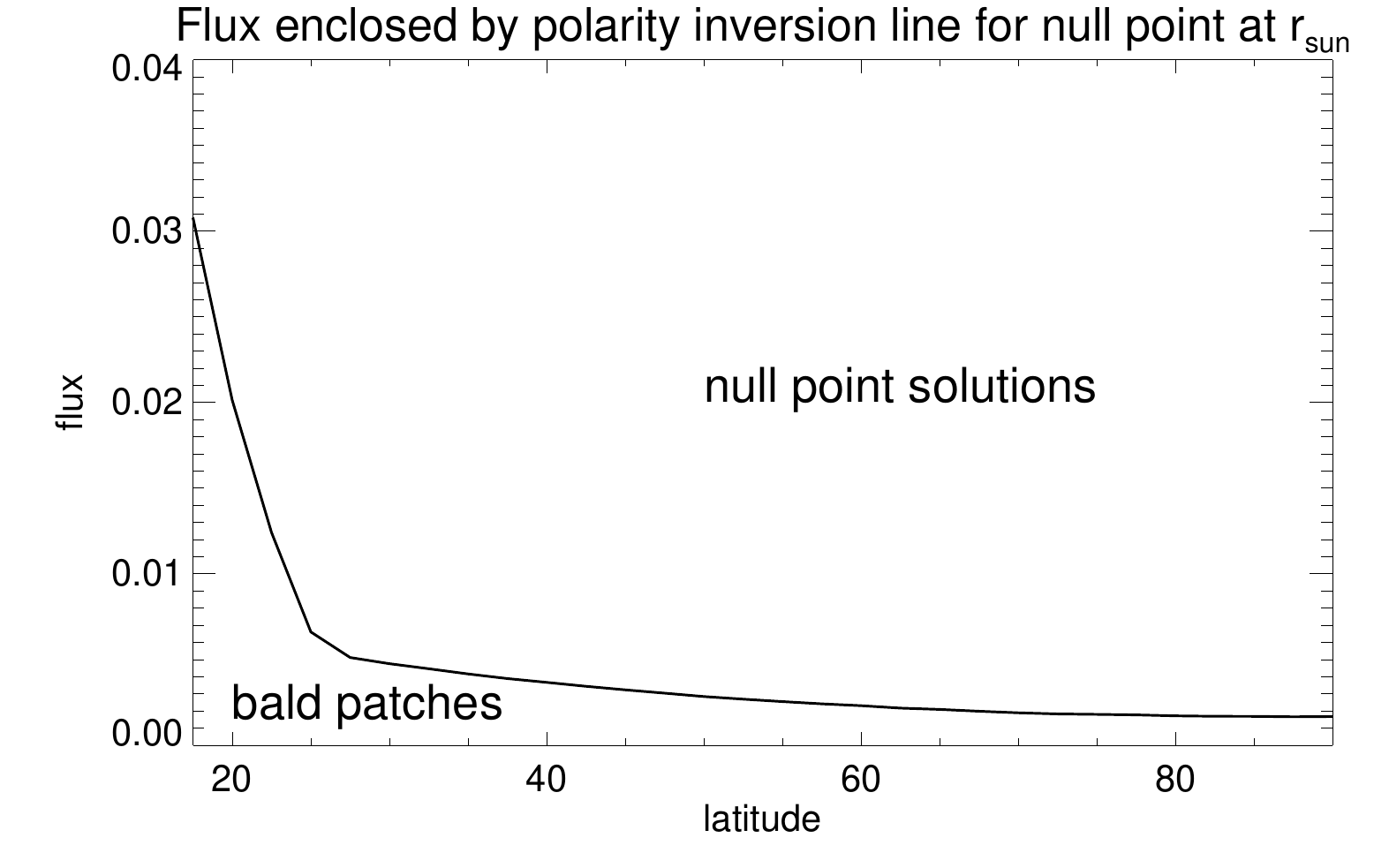}
\end{center}
\caption{\label{fig:flux}Magnetic flux enclosed by PIL as a function of solar latitude for a null point at $r = r_\odot$. The plot shows that the amount of opposite polarity flux in the vicinity of a bald patch decreases strongly with latitude. This means that an embedded dipole with a given normalized surface magnetic flux of e.g.\ 0.01 can only give rise to a bald patch at latitudes of less than about $25^\circ$.}
\end{figure}

\begin{figure}[t]
\vspace*{2mm}
\begin{center}
\includegraphics[width=8.3cm]{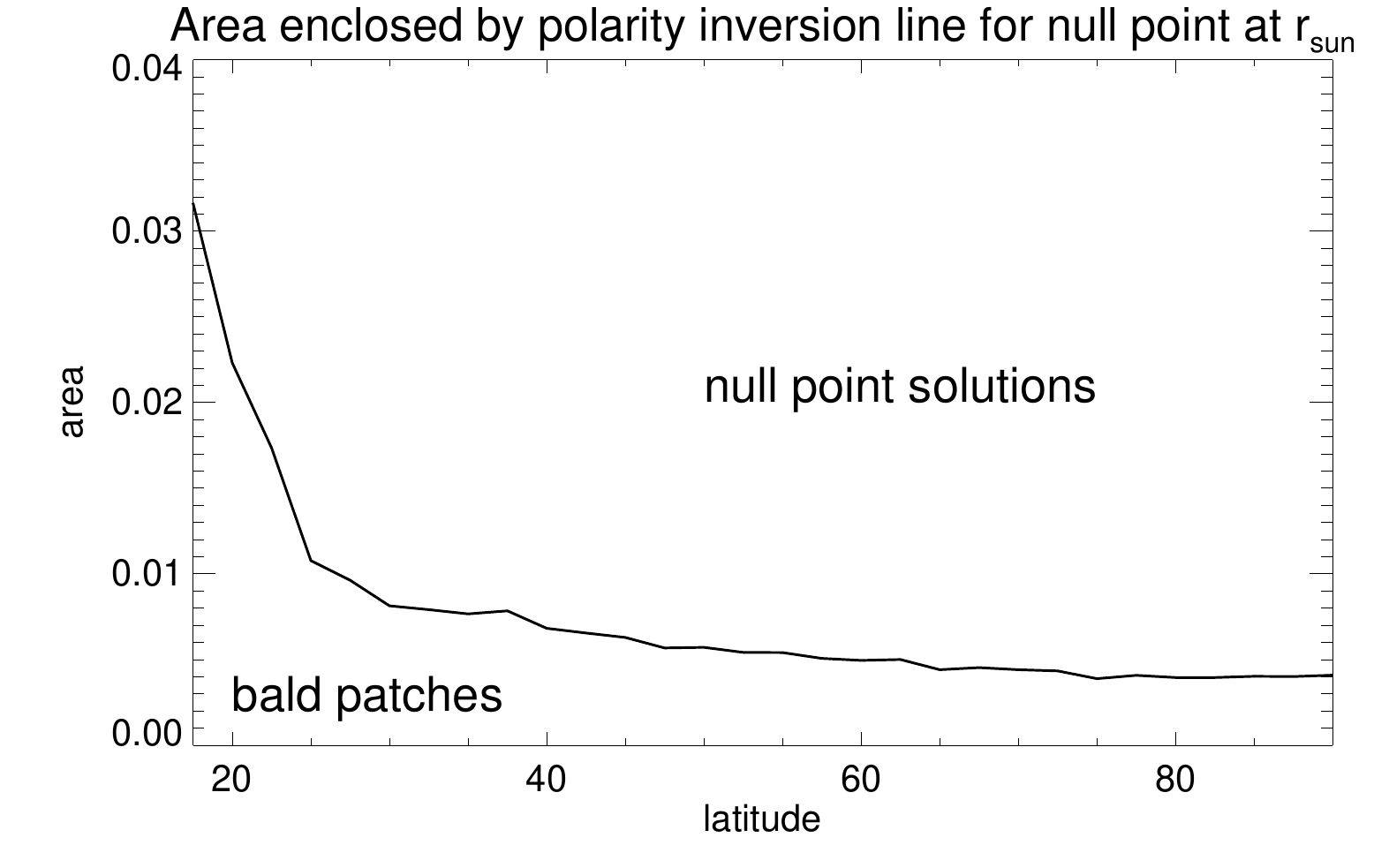}
\end{center}
\caption{\label{fig:area}Area enclosed by polarity inversion line as a function of solar latitude for a null point at $r = r_\odot$. This plot shows that large nested polarities only occur close to the equator.}
\end{figure}

\begin{figure}[t]
\vspace{2mm}
\begin{center}
\includegraphics[width=8.3cm]{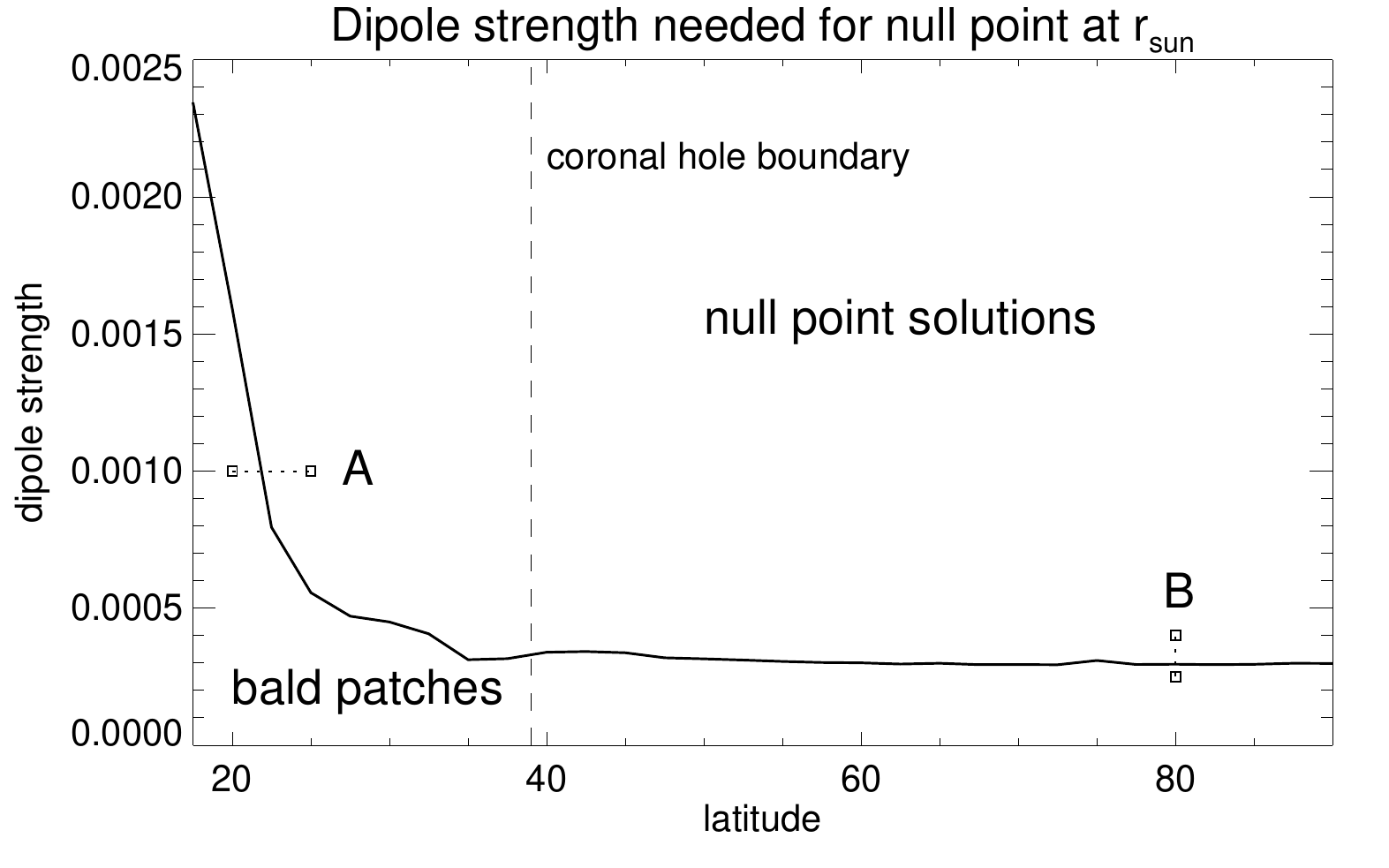}
\end{center}
\caption{\label{fig:dipstrength}Dipole strength needed to produce a magnetic null at the solar surface as a function of solar latitude. The squares connected by a dotted line indicate the two situations described in Sections~\ref{case_A} and \ref{case_B}.}
\end{figure}

Figure \ref{fig:flux} displays the magnetic flux enclosed by the PIL as a function of solar latitude for a null point at $r = r_\odot$. Combinations of latitude and dipole strengths that lie below this curve can support a bald patch topology, while those above will produce a null point above the surface. It can be seen that the amount of opposite polarity flux in the vicinity of a bald patch (i.e.\ the flux contained inside the associated PIL) decreases strongly with latitude. This means that an embedded dipole with a given surface magnetic flux can only give rise to a bald patch at latitudes below a critical value. For a relative flux of e.g.\ 0.01, the critical latitude is about $25^\circ$.
Figure \ref{fig:area} shows the area of these nested polarities as a function of latitude. The shape of this curve is very similar to the one for the magnetic flux which indicates that the mean radial magnetic field inside the nested polarities does not change substantially as a function of latitude.
Finally, Fig.\ \ref{fig:dipstrength} shows the corresponding strength of the embedded dipole that is needed to place the magnetic null exactly at the solar surface. Taken together, these plots show that with increasing latitude, both the flux and size of nested polarities that are associated with bald patches strongly decrease. Already above $25^\circ$ latitude, the parameter regime in which we expect to find bald patches is very small, and especially inside a coronal hole (the approximate location of the coronal hole boundary is indicated in Fig.\ \ref{fig:dipstrength}), we expect most embedded dipole to appear as null points in the corona instead of bald patches.
To verify that the functional dependence found here is indeed the separator between null point and bald patch solutions, we examined two specific cases which are marked "A'' and "B'' in Fig.\ \ref{fig:dipstrength}.

\subsection{\label{case_A}Case A: Constant dipole strength}

\begin{figure*}[ht]
\vspace*{2mm}
\begin{center}
\includegraphics[width=0.49\textwidth]{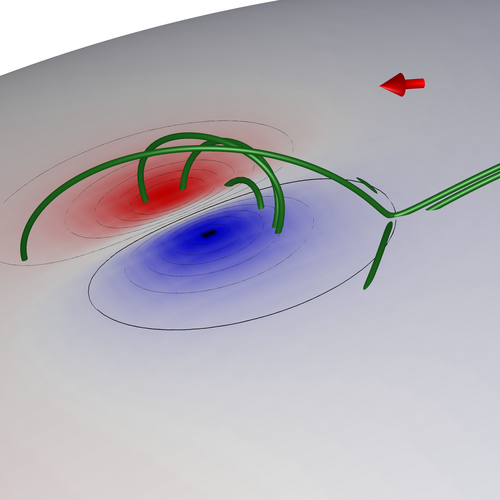}
\includegraphics[width=0.49\textwidth]{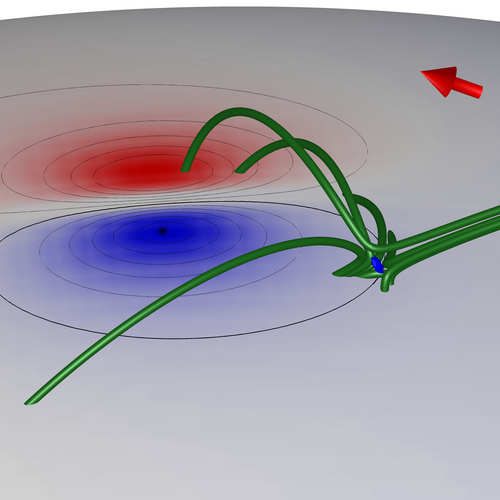}
\end{center}
\caption{\label{fig:lconst}Magnetic field lines around the polarity inversion line for $20^\circ$ (\emph{left}) and $25^\circ$ latitude (\emph{right}). The left plot shows a bald patch topology, while the right one shows a null point topology. The red arrows point north.
}
\vspace*{4mm}
\end{figure*}

 Keeping the dipole strength constant at $l=0.001$, we inspected the field topology on both sides of the separator line. Figure~\ref{fig:lconst} shows magnetic field lines around the PIL for latitudes of $20^\circ$ (\emph{left}) and $25^\circ$ (\emph{right}). As expected, we find a bald patch topology for $20^\circ$ latitude (associated with the little square on the left hand side of the separator line in Fig.\ \ref{fig:dipstrength}) and a null point for $25^\circ$ latitude (associated with the little square on the right hand side of the separator line).

\subsection{\label{case_B}Case B: Constant latitude}

In the second example, we kept the latitude of $80^\circ$ fixed and varied the dipole strength from $2.5\cdot10^{-4}$ to $4\cdot10^{-4}$. 
\begin{figure*}[ht]
\vspace*{2mm}
\begin{center}
\includegraphics[width=0.49\textwidth]{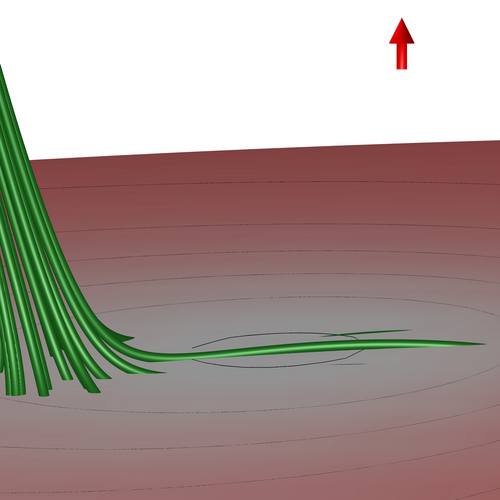}
\includegraphics[width=0.49\textwidth]{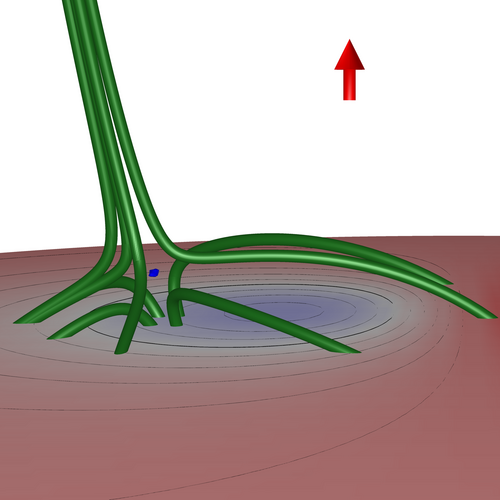}
\end{center}
\caption{\label{fig:thconst}Magnetic field lines around the polarity inversion line at $80^\circ$ latitude (close to the pole) for dipole strengths of $2.5\cdot10^{-4}$ and $4\cdot10^{-4}$. The left plot shows a (very shallow) bald patch topology, while the right one shows a null point topology. The red arrows point north.
}
\vspace*{4mm}
\end{figure*}
Figure~\ref{fig:thconst} shows magnetic field lines around the PIL for these two cases. Even for this high latitude close to the pole, we find a  - very shallow - bald patch topology for the dipole strength that lies below the separator line ($l = 2.5\cdot10^{-4}$), while there is a null point for the dipole strength above the separator line ($l = 4\cdot10^{-4}$). This supports the earlier statement that the function $l_{\rm crit}(\theta)$ which places the magnetic null right at the solar surface is indeed a separatrix between the two regimes. At first sight, the theoretical possibility  of bald patches at and close to the solar pole seems slightly unexpected since one would not commonly associate this open-field region with a habitat for magnetic bald patches. However, Figs.~\ref{fig:flux} and \ref{fig:area} show that the corresponding nested polarity regions are very small and contain so little flux that they may easily be destroyed by the convective motions in the photosphere.

\section{\label{obs}XRT Observations of coronal hole jets}

\begin{figure*}[ht]
\vspace*{2mm}
\begin{center}
\includegraphics[width=\textwidth]{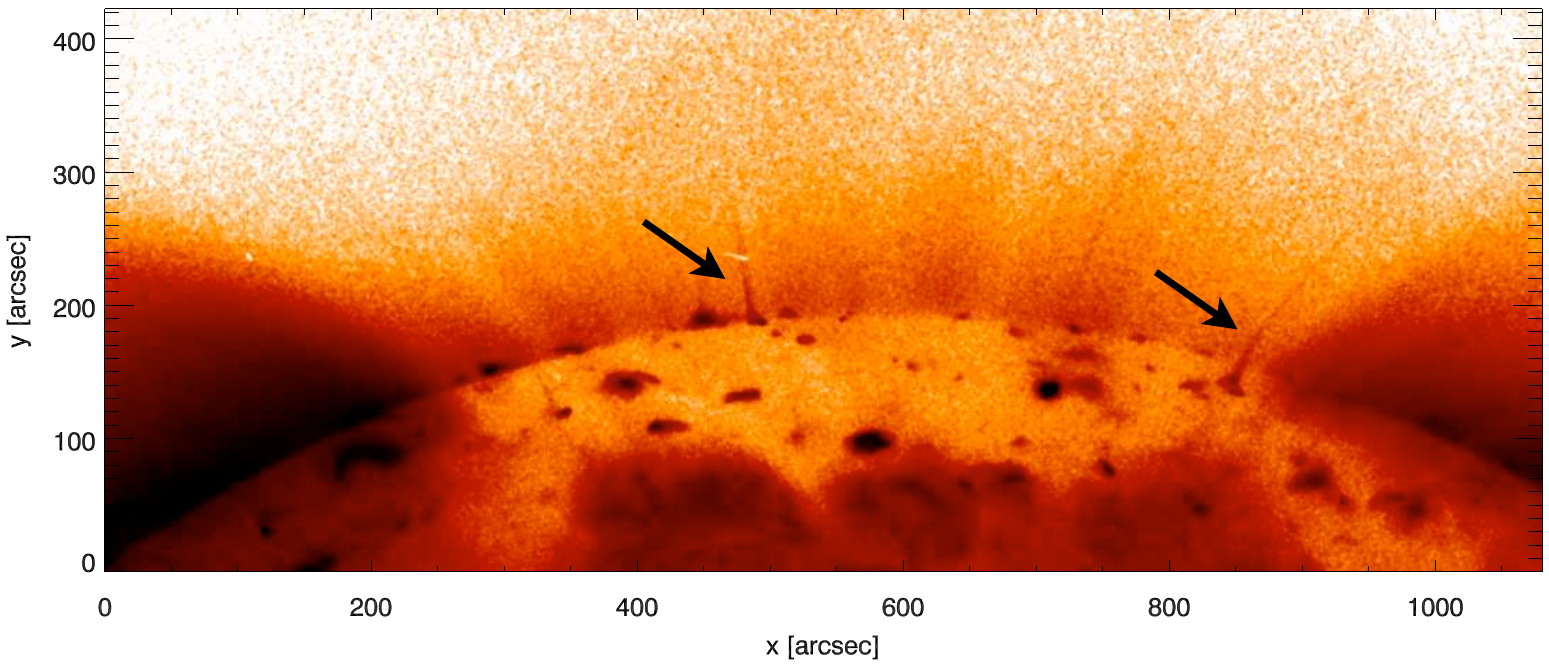}
\end{center}
\caption{\label{fig:xrt}Image taken by the X-Ray Telescope \emph{XRT} on board Hinode on January 10, 2007 using the Al/Poly filter. The dark shading corresponds to high intensity, the light shading to low intensity. The arrows indicate two elongated structures inside the coronal hole which show jetting activity. XRT photo credit: JAXA/NASA/NAOJ/SAO.}
\end{figure*}

Recent observations with the X-Ray Telescope \emph{XRT} on board the recently launched Hinode satellite frequently show coronal hole jets \citep{Cirtain+al2007AAS,Savcheva+al2007AAS}. These jets are very dynamic and relatively short-lived, and especially in their later stage often seem to outline a null point topology with a spine and a fan. We conjecture that these observations may be related to the transition from a bald patch to a null point topology. When the overlying field opens up or becomes sufficiently radial, a bald patch topology will transform into a null point topology. However, this process cannot take place smoothly in ideal MHD unless the submerged null point can emerge through the photosphere. This would mean, however, that concave up field lines that reach several scale heights deep into the convection zone would have to be lifted above the photosphere which seems unlikely given the amount of mass they carry. In these cases, we therefore expect violent coronal dynamics that could e.g.\ give rise to the observed coronal hole jets. Figure~\ref{fig:xrt} shows an image of a northern polar coronal hole, taken with XRT's Al/Poly filter. The dark shading corresponds to high intensity, the light shading to low intensity, and the arrows indicate two elongated structures inside the coronal hole which show jetting activity.
%f

\subsection{\label{comp}Comparison between observations and model}

In order to relate our results to observations, let us compare the fluxes and areas of bright points that show jetting activity. These bright points are typically referred to as "flaring bright points"  and have a broad range of fluxes and lifetimes,  $F = t \cdot 10^{19}$\,Mx $= t \cdot 10^{11}$\,Wb, where $F$ is the flux and $t$ is the lifetime of the bright point measured in days \citep{Golub2008}. The typical size for the radius of such a bright point (which is not a point in the geometrical sense) is on the order of 10\,Mm, which for a circular area corresponds to $A = \pi \cdot 10^{18}$\,cm$^2$. This results in a mean magnetic field of $\langle B_r\rangle =F/A$ of about 3\,G times the lifetime in days. In direct measurements, maximal field strengths of the order of the ones found in plage regions (about 100\,G) are typically detected.

Let us compare these numbers with a bald patch/null point transition at $\theta = 70^\circ$ latitude which in our scenario can trigger a coronal hole jet. Figure \ref{fig:flux} and \ref{fig:area} indicate that the separator between bald patch and null point solutions, around which we expect jets to take place, occurs at a normalized flux of $F  \approx 2 \cdot 10^{-3}$ and a normalized area enclosed by the PIL of $A_{PIL} = 4 \cdot 10^{-3}$. This results in a mean magnetic field of $\langle B_r \rangle = 1/2 \cdot B_{r,{\rm pole}}$. In physical units, this area corresponds to $A = 4 \cdot 10^{-3} \cdot A_\odot/ 4\pi \approx 2 \cdot 10^{19}$\,cm$^2$, i.e. a circle with a radius of 25\,Mm. This area is about a factor of 6 larger than the area of a typical bright point, but on the other hand the field strength is weaker than what is observed.
Taking the radial field strength at the solar pole of our model, $B_{r, {\rm pole}}$ = 10\,G, the resulting flux enclosed by the PIL is $F = 9.6 \cdot 10^{19}$\,Mx, which is of the same order of magnitude as the observed flux for bright points that live longer than a day. Placing the embedded dipole slightly deeper below the surface in the model would decrease the area enclosed by the PIL and thus the flux, bringing it even closer to the observed estimates.
We therefore conclude that the parameter regime of magnetic fluxes and bright point areas studied in this work is in agreement with the range of observed values and thus is realistic. We expect that the Solar Dynamics Observatory (SDO), with its ultra-high spatial and temporal cadence, will be able to test our flaring bright point scenario with great precision.

\conclusions[\label{discussion}Discussion]

Using a simple magnetostatic model, we have shown that magnetic bald patches are more likely to occur in closed-field regions at low latitudes than at high latitudes, especially in coronal holes. While there is no unique size range for bald patches, the larger the bald patch is, the less likely it is to occur at high latitude.
If bald patches encounter open-field surroundings, either by moving into a coronal hole or by the opening-up of the surrounding field, the bald patch topology almost invariably transforms into a null point topology with a spine and a fan. The time-dependent evolution of this scenario will be very dynamic since the change from a bald patch to null point topology cannot occur via a simple ideal MHD evolution in the corona. Even with magnetic reconnection, it is not clear how this process will take place. One event that needs to occur is that a null point must form in the corona, where there was not one before. Second, the concave up field along the bald patch position of the PIL must transform into concave down field.  It is not obvious how current sheet formation and magnetic reconnection can accomplish this. A dynamic calculation that includes some of the photospheric region is clearly needed.
This mechanism of topologically driven dynamics may be important to understand the evolution of e.g.\ coronal hole jets such as those observed by Hinode's XRT instrument. Time-dependent 2D and 3D MHD simulations are planned to fully assess this scenario.

\begin{acknowledgements}
This work was supported in part by NASA, ONR and the NSF-SHINE program. Hinode is a Japanese mission developed and launched by ISAS/JAXA, with NAOJ as domestic partner and NASA and STFC (UK) as international partners. It is operated by these agencies in co-operation with ESA and NSC (Norway). We thank the anonymous referees for helpful comments that improved the quality of the manuscript.
\end{acknowledgements}

\bibliographystyle{copernicus}
\bibliography{aamnem99d,MHD}
%%

%\addtocounter{figure}{-1}\renewcommand{\thefigure}{\arabic{figure}a}

\end{document}